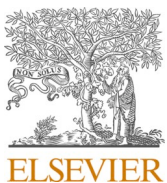

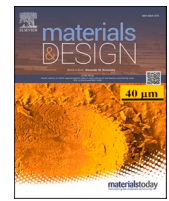

# Novel nanoindentation strain rate sweep method for continuously investigating the strain rate sensitivity of materials at the nanoscale

Hendrik Holz [a, b, c, *], Benoit Merle [a, b]

[a] Institute of Materials Engineering, University of Kassel, Mönchebergstr. 3, 34125 Kassel, Germany
[b] Materials Science & Engineering, Institute I, Friedrich-Alexander-Universität Erlangen-Nürnberg, Martensstr. 5, 91058 Erlangen, Germany
[c] Max-Planck-Institut für Eisenforschung GmbH, Max-Planck-Straße 1, 40237 Düsseldorf, Germany

ABSTRACT

We introduce a new nanoindentation method to continuously measure the hardness while sweeping through orders of magnitudes of strain rates within a single experiment. While nanoindentation already allows the determination of the strain rate sensitivity of materials by means of strain rate jump tests, these are typically limited to few discrete strain rates. With the new method, the strain rate sensitivity can be measured continuously as a function of the strain rate. Applications to fused silica, Zn-22 %Al superplastic alloy, single crystalline aluminum, various nanocrystalline metals and a palladium-based metallic glass are shown. Besides some discrepancy with the reference measurements, the new method seems only affected by the presence of a strong nanoindentation size effect. Provided this indentation size effect is not excessively large and can be corrected for accurately, the method proves robust, with no suggestion that the direction of the strain rate sweep affects the evaluation of the strain rate sensitivity.

## 1. Introduction

The mechanical strength of materials very often depends upon the applied strain rate. This needs to be considered for obtaining reliable mechanical assessments, e.g. about the material behavior during high rate forming processes. The knowledge of the strain rate sensitivity (SRS) and the hardness of these materials at specific strain rates are hereby paramount. Such characterization can be performed by nanoindentation, with strain rate defined as [1]:

$$\dot{\varepsilon} = \frac{\dot{h}}{h} = \frac{1}{2}\left(\frac{\dot{P}}{P} - \frac{\dot{H}}{H}\right) \approx \frac{\dot{P}}{2P} \qquad (1)$$

where $h$ is the indentation depth, $P$ is the applied load, $H$ is the hardness and $\dot{h}$, $\dot{P}$ and $\dot{H}$ are the respective time-derivatives. However, as Leitner et al. showed, the strain rate is only constant during the loading segment of constant $\dot{P}/P$ experiments [2], therefore it is best to combine constant strain rate (CSR) indentation with the continuous stiffness method (CSM), rather than performing quasi-static nanoindentation. To obtain the strain rate sensitivity (SRS), several experiments at different CSR can be compared. The SRS, usually termed $m$, can then be determined by

analyzing the slope of the experimental hardness data in an inverse Norton plot [3]:

$$m = \frac{d(lnH)}{d(ln\dot{\varepsilon})} \qquad (2)$$

Finally, the SRS can be used to calculate the activation volume $V$ as [4]:

$$V \cong c\sqrt{3}\,\frac{kT}{mH} \qquad (3)$$

with k being the Boltzmann constant and $T$ the absolute temperature. In this equation, the hardness $H$ is related to the flow stress by a constraint factor $c$ which is usually close to 3 [5].

A major concern with CSR testing at slow rates is that the experiments take several minutes up to half-an-hour, so that the data are strongly affected by thermal drift [4]. To reduce the overall testing time for accessing lower strain rates, Maier et al. and Alkorta et al. independently implemented sudden variations of the strain rate during a single experiment [4,6]. The so-called strain rate jump tests (SRJ) are now widely used for measuring the SRS [7,8]. Besides pointed nanoindentation testing, SRJ tests are also used to study the SRS by micropillar compression testing [9,10]. While thermal drift limits the lowest achievable strain rates for nanoindentation, the plasticity error limits CSM measurements at high strain rates [11,12]. The Merle-Higgins-Pharr method allows to bypass the plasticity error and can be used to test strain rates in the dynamic range in a controlled manner, but so far only tests at CSR were performed at high strain rates [13] and SRJ tests in the intermediate regime [14].

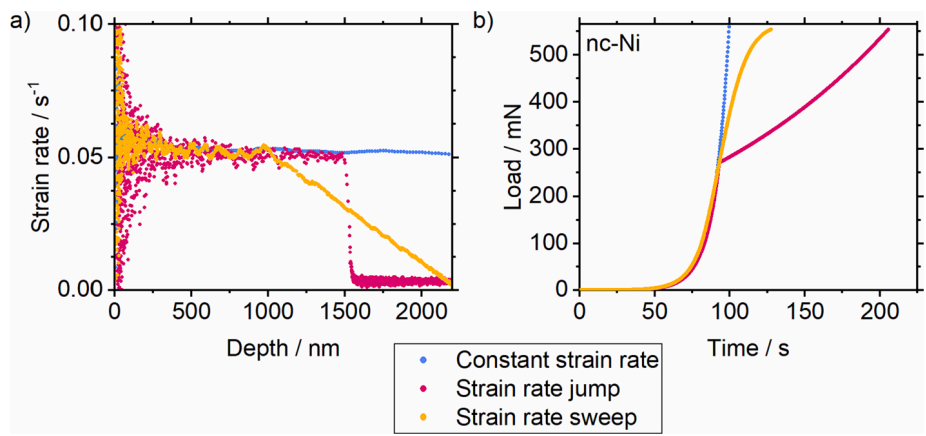

**Fig. 1.** a) strain rate profiles of CSR, SRJ and the newly developed SR-sweep method. b) Corresponding load profiles. The figure showcases nc-Ni, but the profiles are similar for other materials. (For interpretation of the references to color in this figure legend, the reader is referred to the web version of this article.)

Unlike CSR, further nanoindentation techniques intrinsically involve a variation of the applied strain rate during a measurement. In creep experiments, the load is held constant while the hardness is continuously measured over decaying strain rates [15,16]. To avoid the influence of drift, the contact stiffness from CSM is used to evaluate the displacement independently from thermal drift effects [15]. Similarly, during impact tests, the hardness is continuously measured while the applied strain rate decays from ballistic values [17–20]. In these two techniques, the variation of the strain rate is rather uncontrolled and unpredictable.

In the next sections, we introduce a novel method for nanoindentation, which enables controlled measurements of the hardness and SRS continuously across a wide range of strain rates.

## 2. Strain rate sweep method

Nanoindentation strain rate sweeps are performed by linearly varying the indentation strain rate over the indentation depth during a CSM experiment, as opposed to an abrupt change in strain rate in SRJ tests [4]. A typical experiment starts with a CSR loading segment, which is not evaluated and only helps overcome the shallow depth range associated with uncertainties in the tip area function and strongest indentation size effect. We use a rather fast initial loading rate of 0.05 s$^{-1}$ in order to reduce the duration of the experiments and alleviate thermal drift issues. The sweep segment is initiated around 1000 nm indentation depth in Fig. 1 a) – see orange curve.

From the measured hardness and strain rate, an inverse Norton plot is then constructed from the measured hardness and strain rate. The hardness is continuously measured from CSM and evaluated according to Oliver and Pharr [21,22]. The strain rate is continuously evaluated from $\dot{h}/h$. We favored this definition over load-based $\dot{P}/2P$, since the latter equation is only an approximation [2] which shows some discrepancies (transients) for highly strain rate sensitive materials such as superplastic alloys. $\dot{h}/h$ was evaluated by Savitzky–Golay quadratic fitting of the displacement data with a custom Python script using a sliding window of 6 data points. Time-linear sweeping of the strain rate leads to equidistant datapoints in the logarithmic Norton plot. Ultimately, the SRS is calculated as the local slope of the Norton curves (see next section).

In Fig. 1, the loading profiles and resulting strain rates of a sweep test are compared to a SRJ and a CSR experiment. Unlike the latter, the loading-time profile of sweep experiments resembles a sigmoid function, see Fig. 1 b).

## 3. Materials and methods

The tested samples, together with data on their origin, Young's modulus to hardness ratio and Burgers vectors, are presented in Table 1. Except for the FS and Sx-Al samples, which were supplied with a smooth and polished surface, all samples were ground and subsequently polished with diamond suspension down to a particle size of 1 μm. A final polishing step with oxide polishing suspension ensured a scratch free surface.

The coupling between indentation depth and applied strain rate makes the SR-sweep method highly sensitive to the indentation size effect (ISE). Therefore, a correction of the recorded hardness according to Nix and Gao [23] was introduced. To this aim, the initial CSR segment was fitted with the Nix-Gao model to determine the indentation length scale $h^*$. By rearranging the Nix-Gao model the following equation was used to calculate the indentation size effect corrected hardness $H_0$:

$$H_0 = \frac{H}{\sqrt{1 + \frac{h^*}{h}}} \tag{6}$$

The hardness values used for fitting the data to the Nix-Gao model were taken between 350 and 1000 nm. As a matter of fact, the data at shallower indentation depth are known to deviate from Nix-Gao [24], while the CSR segment ends at 1000 nm. To reduce the influence of experimental noise, $h^*$ was obtained from data aggregated from multiple tests performed with the same parameters. No significant influence of the different strain rates on the shape of the Nix-Gao correction is expected, as recently shown by Liang and Pharr [25].

**Table 1**
Tested materials, with data on origin, Young's modulus-to-hardness ratio and Burgers vectors.

| Material | Abbrev. | Origin | E/H ratio | Burgers vector |
|---|---|---|---|---|
| Fused silica | FS | Reference sample of G200 | 7 | – |
| Pd based metallic glass | Pd-MG | Pd77.5-Cu6-Si16.5 | 19 | – |
| Nanocrystalline nickel | nc-Ni | Pulsed electron deposition | 33 | 2.5 Å |
| Fe-21 % Cr | nc-FeCr | Fe- 21.1 at% Cr; High pressure torsion | 54 | 2.5 Å |
| Perlitic steel | Steel | Fe- 3.3 at% C | 64 | 2.3 Å |
| Zn-22 % Al | ZnAl | Heat treated and water quenched; Grain size 3.09 μm | 73 | 2.7 Å |
| Single crystalline aluminum | Sx-Al | Reference sample of G200 | 236 | 2.9 Å |

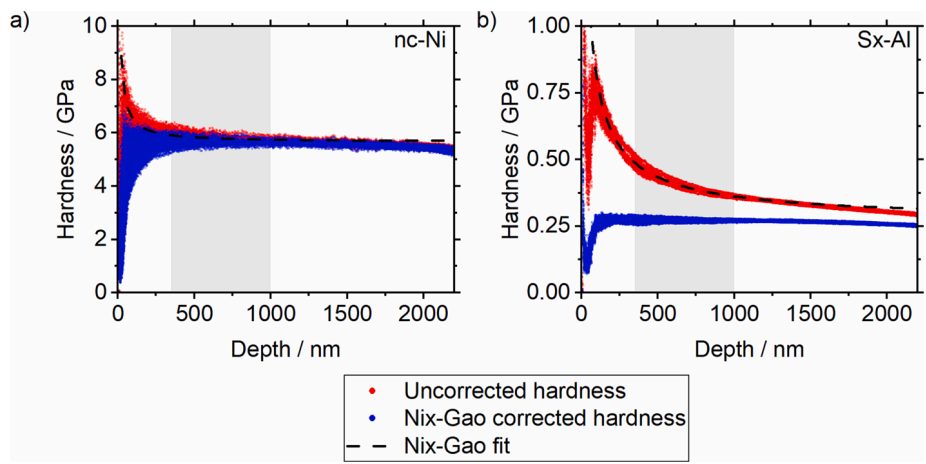

**Fig. 2.** Effect of the Nix-Gao indentation size effect correction on the hardness data obtained from SR-sweep experiments on a) nc-Ni and b) Sx-Al. The correction appears severe for Sx-Al. The grey area indicates the regime used to perform the Nix-Gao fit. (For interpretation of the references to color in this figure legend, the reader is referred to the web version of this article.)

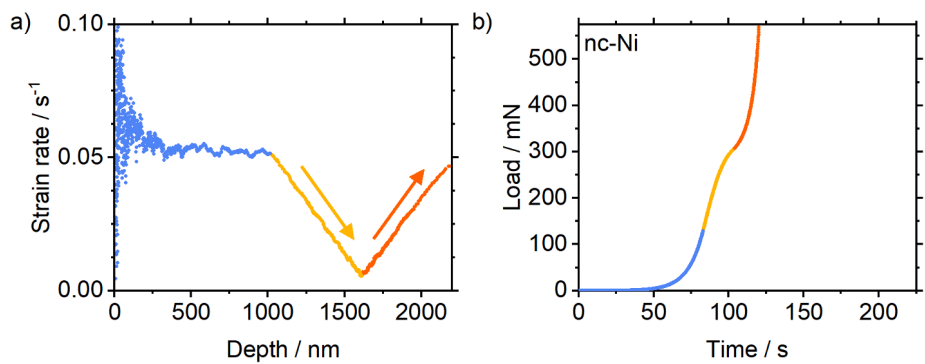

**Fig. 3.** a) strain rate profiles of strain rate sweep reversal experiments. b) corresponding load profile. The figure showcases nc-Ni, but the profiles are similar for other materials. The sweep segment from high to low strain rates is marked in yellow and the segment from low to high strain rates in orange. (For interpretation of the references to color in this figure legend, the reader is referred to the web version of this article.)

Except for the amorphous FS, all samples were corrected with this Nix-Gao procedure. The correction is illustrated in Fig. 2 for sweep experiments on nc-Ni and Sx-Al. While the ISE has only minor influence on the nc-Ni data, it has a severe influence on Sx-Al. In this case, a massive correction of the data is performed, with detrimental consequences to be shown further down. For consistency purposes, the Nix-Gao correction was equally applied to the new SR-sweep and the reference SRJ and CSR methods (see below).

For the steel and ZnAl samples, a strong scatter in the data was found which is likely due to coarser microstructure, different phases and different grain orientations. Since the purpose of this work is to develop a new method and not to investigate the different phases of these materials, statistical outliers were removed and the evaluation focused on the most consistent data.

All experiments were performed on a load-controlled G200 nanoindenter from KLA (Milpitas, CA, USA) using a XP head equipped with a modified Berkovich tip produced by Synton MDP (Nidau, Switzerland). The frame stiffness and tip area were calibrated on fused silica according to the Oliver-Pharr method [21,22]. A CSM frequency of 45 Hz with an amplitude of 2 nm was used for all tests.

The sweep tests were typically performed from the highest ($0.05\ s^{-1}$) to the lowest ($0.001\ s^{-1}$) strain rate. The resulting profiles are shown in Fig. 1 a). At least 6 sweep experiments were evaluated for each material. The main reason for favoring this sweep direction was that it minimizes the testing time, constraining it to <350 s. This reduces thermal drift issues. Additionally, higher strain rates are hardly susceptible to the CSM plasticity error at shallow depth [11,12], and the measured phase angle indeed rarely exceeded 10°, and if so apparently only due to data scattering.

In a further step, we appended a segment with reversed direction to the standard test, in order to evaluate a possible influence of a strain rate history on strain rate sweep tests. A strain rate profile of a sweep reversal experiment is shown in Fig. 3. At least 10 experiments were analyzed for each material. As expected, the measured Young´s modulus remains rather constant during the strain rate sweep, see Fig. S-1 in the Supplementary Information.

For reference measurements of the SRS, constant strain rate CSM measurements were performed in the range $0.05\ s^{-1} - 0.01\ s^{-1}$; or $0.1\ s^{-1} - 0.01\ s^{-1}$ for FS, which is less prone to the CSM plasticity error [11]. The reference values at lower strain rates were obtained from SRJ owing to the requirement to keep the duration below 350 s. To this aim, an initial strain rate segment of $0.05\ s^{-1}$ was followed by one jump to a lower strain rate of $0.01\ s^{-1}$, $0.00316\ s^{-1}$ or $0.001\ s^{-1}$. From these tests, the hardness was evaluated from the depth range where no change in hardness was evident anymore, i.e. between 1500 and 2100 nm to 2200 nm.

The SRS was evaluated differently for the sweep tests and the reference measurements. For the reference measurements, the SRS was obtained from a global linear regression of the inverse Norton plot data. In contrast, the sweep experiments were evaluated in a more continuous manner. The data of the sweep segments were scanned with a sliding window and SRS obtained by linear regression within an interval of 0.5 logarithmic strain rate centered on the current data point. Only intervals including at least 250 data points were analyzed. The standard deviation was taken as the standard error of the slope of the linear fit. For the sweep reversal tests, the two sweep segments were split up and treated independently from each other.

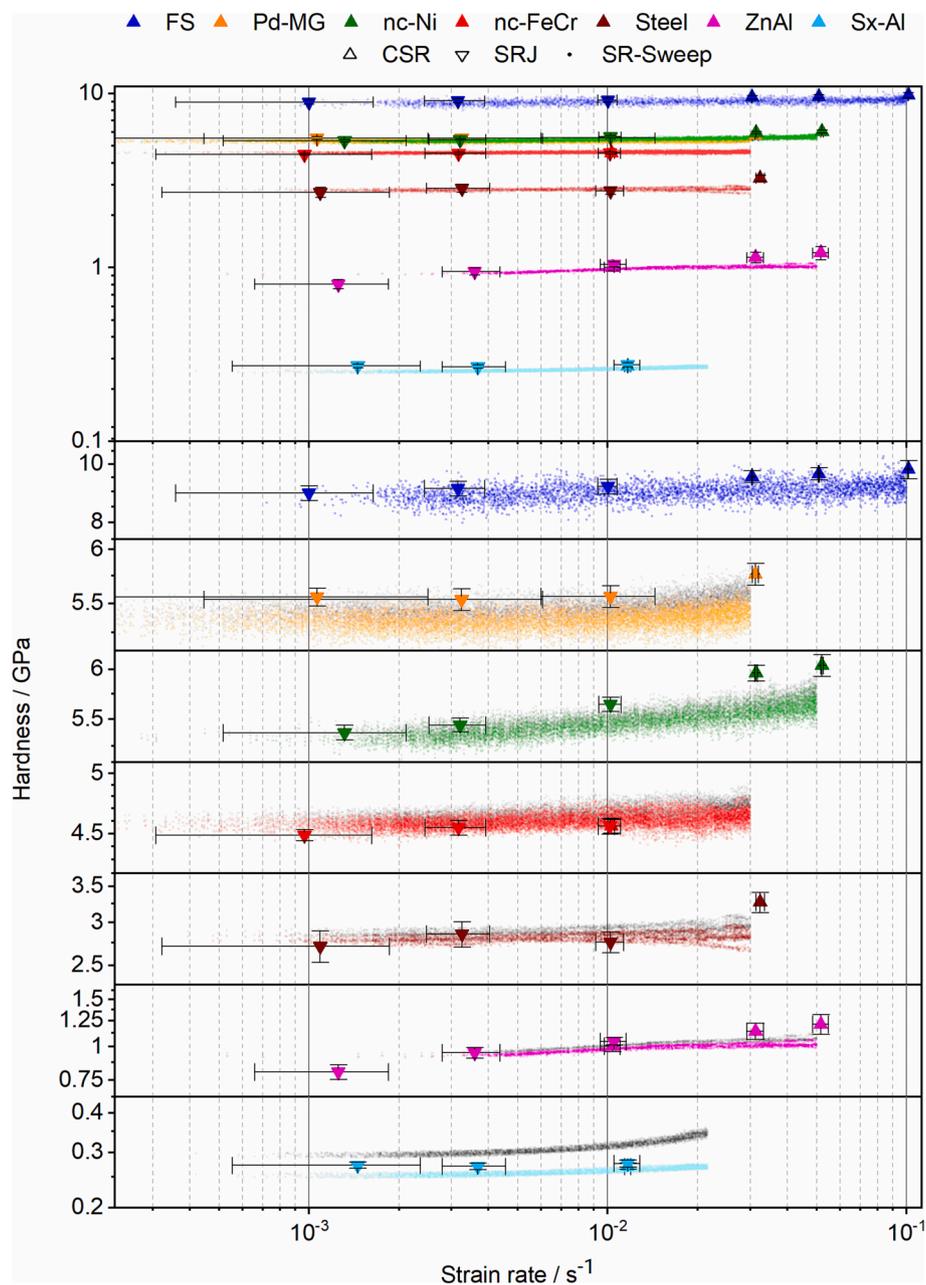

**Fig. 4.** Inverse Norton plot for all tested materials. Each datapoint from the CSR and SRJ tests is averaged from 10 to 16 tests. The data points from the SR-sweep tests originate from 6 to 16 experiments per material. The grey point clouds correspond to the raw data while the colored ones have been corrected for ISE and are considered in the following. (For interpretation of the references to color in this figure legend, the reader is referred to the web version of this article.)

For the evaluation of the activation volume, the Burgers vectors of the different materials were taken from following sources [26–30].

## 4. Results & discussion

### 4.1. Strain rate sensitivity from sweep experiments

The hardness measurements from the sweep experiments are compared to reference methods – CSR and SRJ testing. The results are summarized in the inverse Norton plot shown in Fig. 4, which includes only valid tests, i.e. with a phase angle < 10°. The overall good agreement is a first step towards validating the new SR-sweep test technique.

It can be seen that the experimentally probed strain rate range does not perfectly align between SR-sweep and reference measurements. On the one hand some sweep tests end at higher strain rates than the reference measurements. In case of ZnAl, this is likely due to extensive creep, which superimposes to the targeted strain rate. In the case of fused silica, the reason is likely the feedback control frequency of the indenter, which cannot keep up with the fast changes of the strain rate. This results in the strain rate lagging behind. Therefore, the final indentation depth is reached and the experiment stops before the lowest strain rate target is reached. These issues could be avoided with a displacement-controlled nanoindenter with fast electronics. Another way to achieve measurements at lower strain rates could be to implement a stiffness-based depth sensing similar to [15]. On the other hand, with nc–FeCr the SR-sweep could start from a higher strain rate than the reference CSR measurements. This is because the former data were collected from a shallower range than the latter ones. As a result, the SR-sweep data are unaffected by the CSM plasticity error, while the corresponding CSR data exhibited a phase angle > 10° and had to be discarded.

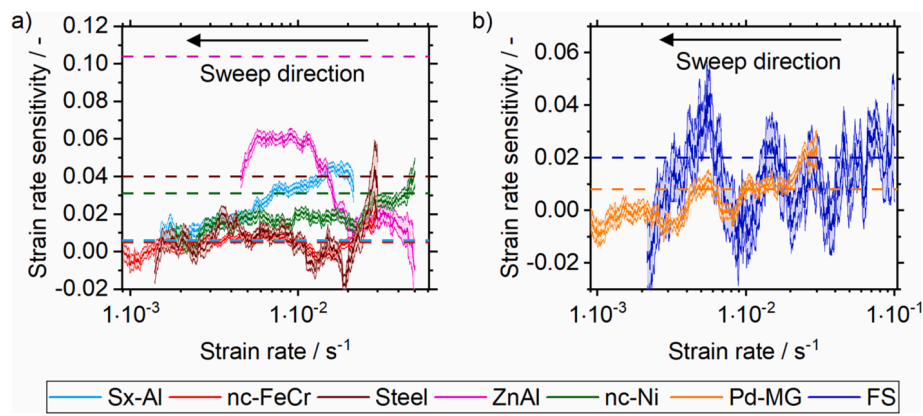

**Fig. 5.** Strain rate sensitivity as a function of strain rate. The dashed lines represent the SRS obtained from a linear regression of the CSR and SRJ data in the inverse Norton plot, while the solid lines represent the results from the SR-sweep method. a) SRS of crystalline materials. b) SRS of amorphous samples. The presented data were averaged from 6 to 16 experiments and the error bars represent the standard deviation of the linear term from the regression. (For interpretation of the references to color in this figure legend, the reader is referred to the web version of this article.)

All strain rate sensitivities calculated from the SR-sweep tests are shown in Fig. 5 a) for crystalline materials and in b) for amorphous materials. The dotted horizontal lines correspond to the SRS from the reference measurements and the continuous lines are calculated from the SR-sweep experiments.

While the measured SRS is very well in line with the reference value for nc–FeCr, the other samples show a divergence, which is largest in the case of ZnAl, see Fig. 5. There does not appear to be any systematic trend, with SR-sweep tests showing either a larger or a smaller SRS than reference measurements. The origin of the deviation remains unclear, but might be related to an artefact in the evaluation, or a strain rate history dependence. For most samples, the activation volume globally decreases with increasing strain rate, see Supplementary Information Fig. S2. This is a sensible trend, meeting the expectations for thermally activated deformation. Additionally, the SRS decreases with decreasing strain rate only in the beginning of the SR-sweep experiment. Again, it is unclear what caused this initial decrease of the strain rate. It might be related to feed-back control from the machine, artefacts from the data analysis or a strain rate history dependence. Nevertheless, the SRS evaluated with the SR–sweep method are broadly in line with literature values of Sx-Al [10], nc-Ni [4,9] and FS [2].

The superplastic alloy Zn-Al is the only sample not showing a monotonic trend of SRS after the initial decrease in SRS. This was expected from previous publications by Feldner et al., which predicted that regions 2 and 3 of the superplastic deformation behavior would be found at room temperature for such a fine microstructure [8].

FS, shown in Fig. 5b, evidences a high scatter in SR-sweep data. The SRS of FS varies unexpectedly between −0.03 and 0.05 and does not exhibit any meaningful trend. The mean value of 0.01 is slightly lower than our reference measurements. This scatter in SRS originates from noise in the stiffness data, which propagates to the evaluation of the hardness. This noise appears stronger for FS than for any other tested material. In the future, this issue could be mitigated by filtering the data with a low-pass filter or by improving the CSM data collection, e.g. using the approach developed by Phani et al. [31].

Generally, the measurements evidence an improved resolution in terms of strain rate in comparison to the state of the art. This makes the new method attractive for materials whose SRS is strongly dependent on the strain rate, e.g. superplastic alloys [8]. The continuous measurement of the activation volume can additionally lead to a better assessment of brittle-to-ductile transitions at the nano-scale, which temperatures also depend on the strain rate [32,33].

### 4.2. Strain rate history dependence of SRS

In order to investigate a possible influence of the strain rate history dependence on the SRS, reverse SR-sweep tests were introduced, as previously described in Fig. 3 a). The resulting SRS values are shown in Fig. 6. The yellow data corresponds to the first SR-sweep regime going from high to low strain rates, while the orange data is from the opposite SR-sweep direction (slow to fast).

FS and Pd-MG show a high scatter in both sweeping directions over the investigated strain rate range. This is respectively caused by a high experimental noise in the CSM signal of FS, and the serrated flow of Pd-MG.

There is a strong discrepancy between sweeping directions in the case of Sx-Al. It most likely originates from the large ISE, which is stronger for Sx-Al than for any other investigated material. Although we strove to mitigate the indentation size effect with a Nix-Gao based correction, it appears insufficiently accurate, and it remains challenging to decouple strain rates from the indentation depth under such extreme conditions.

For all other samples, we observe a good agreement between the SRS measurements performed in both sweeping directions. The only exception is the beginning of the measurement, i.e. the initial portion of the first SR-sweep segment (from high to low SR). The corresponding SRS data appear out of line, as was already observed in Fig. 5.

Notwithstanding this initial artefact, in case of the superplastic alloy ZnAl the different regimes of deformation are clearly visible with a SRS close to 0 at the lowest SR followed by an increase of the SRS until a steady state is reached at the highest SR.

In addition, sweep experiments with varied initial and final strain rates were performed on the nc-Ni and nc-FeCr samples, see Fig. 7. The measured SRS values overlap quite well, which suggests that the new SR-sweep method is not susceptible to the SR slew rate.

All in all, there is no suggestion of strain rate history affecting the measurements, provided the indentation size effect has been accurately corrected for.

## 5. Conclusions

A new method to measure the strain rate sensitivity continuously over a broad range of strain rates was presented. Several materials with different microstructures and mechanical properties were investigated, and the measured strain rate sensitivity was compared to reference measurements from well-established methods. Although the new method has a tremendous potential, we identified two current limitations.

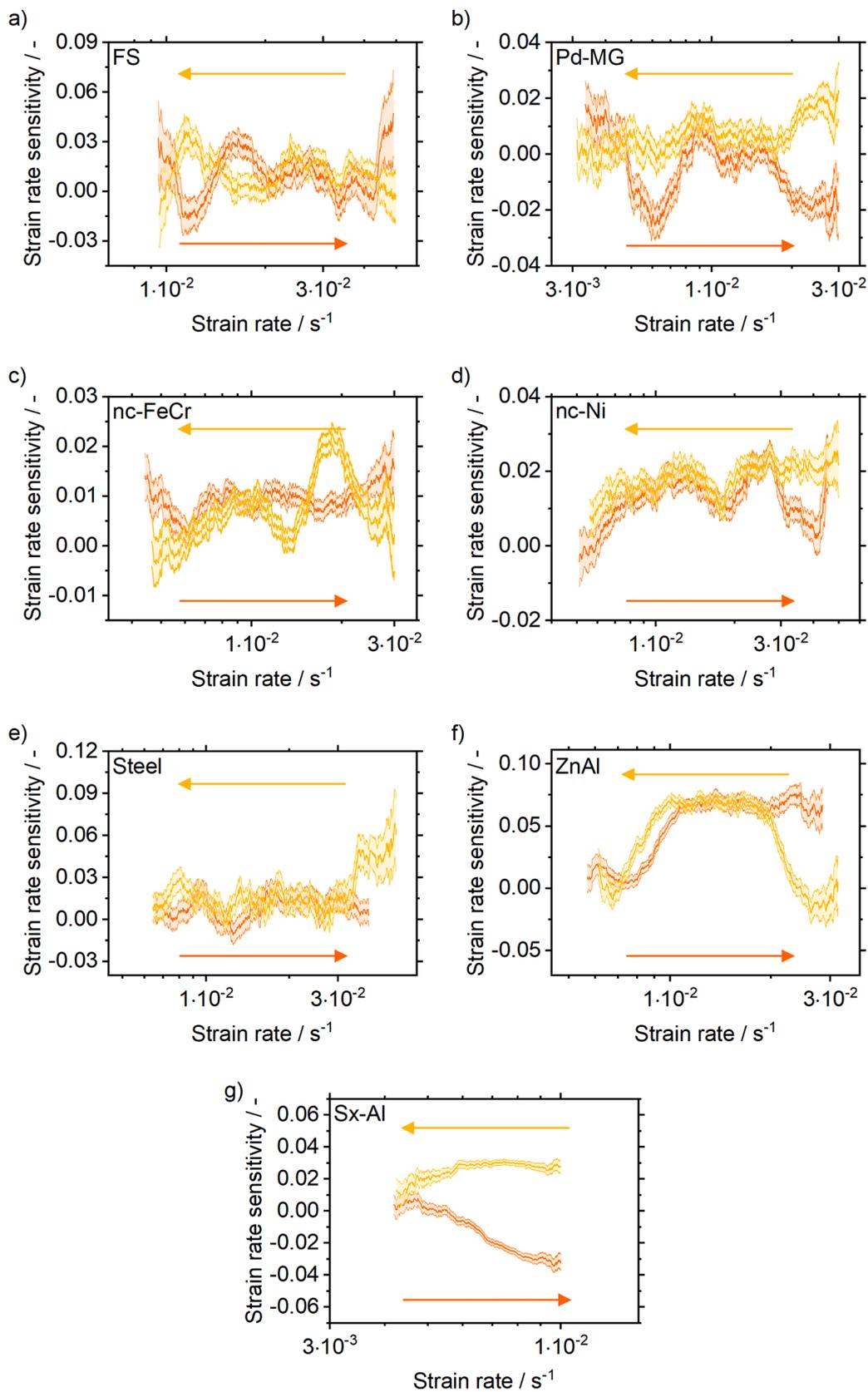

**Fig. 6.** Strain rate sensitivity from reversed SR-sweep experiments. a) FS. b) Pd-MG. c) nc-Ni. d) nc-FeCr. e) Steel. f) ZnAl. g) Sx-Al. The first SR-sweep (yellow curve) is from high to low strain rates and the follow-up reversed direction (orange curve) from low to high strain rates as indicated by the arrows. (For interpretation of the references to color in this figure legend, the reader is referred to the web version of this article.)

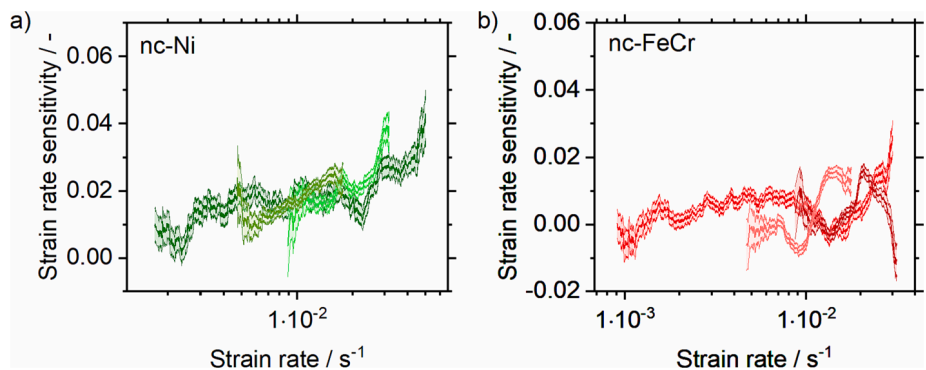

**Fig. 7.** SRS from Sweep experiments over different sweep ranges. a) nc-Ni. b) nc-FeCr. (For interpretation of the references to color in this figure legend, the reader is referred to the web version of this article.)

On the one hand, some behaviors observed during the SR-sweep measurements would need further investigation. Firstly, it should be noted that the agreement with the reference measurements by SRJ and CSR testing is far from ideal. Secondly, the SRS evaluated at the beginning of the SR-sweep regime often falls out of line, compared to the rest of the measurement. This could be a consequence of machine effects, data post-processing, or a strain rate history dependence of the material.

On the other hand, the coupling between strain rate and indentation depth during the sweep makes the new method sensitive to indentation size effect. Performing a perfect correction of the ISE is paramount. This became obvious for the Sx-Al sample exhibiting a strong indentation size effect. The indentation size effect prevents an application to soft, coarse-grained materials. In future, this issue could be addressed by performing more advanced ISE correction techniques such as [34].

There is potential to optimize the novel SR-sweep method. For instance, the scattering in SRS could be reduced by using advanced filtering or by using better CSM implementations to measure the contact stiffness. Additionally, a wider range of strain rates can be assessed by enhancing the instrumentation, so as to access lower and higher strain rates. Higher strain rates are accessible with nanoindenters supporting faster CSM frequencies, therefore reducing the phase angle and in turn avoiding the plasticity error [12] or by using the Merle-Higgins-Pharr method to evaluate the hardness [13].

### Declaration of Competing Interest

The authors declare that they have no known competing financial interests or personal relationships that could have appeared to influence the work reported in this paper.

### Data availability

The experimental data and G200 method files are available on Zenodo and can be downloaded from: https://doi.org/10.5281/zenodo.10033257.

### Acknowledgements

The authors want to thank M. J. Duarte for providing many of the tested samples and M. J. Duarte, J. Wheeler and J. P. Best for helpful discussions, M. J. Duarte and R. Ramachandramoorthy for reviewing a draft of the paper as well as G. Dehm for the opportunity to perform measurements at the Max-Planck-Institut für Eisenforschung GmbH and M. Göken for allowing to conduct further measurements at the Institute I, Materials Science & Engineering department at Friedrich-Alexander-Universität Erlangen-Nürnberg after moving to University of Kassel. This project has received funding from the European Research Council (ERC) under the European Union's Horizon 2020 research and innovation programme (grant agreement No. 949626). Open access funding enabled and organized by Projekt DEAL.

### Appendix A. Supplementary data

Supplementary data to this article can be found online at https://doi.org/10.1016/j.matdes.2023.112471.